\documentstyle[12pt]{article}

\newlength{\extraspace}
\newlength{\extraspaces}
\newcommand{\be}{\begin{equation}
\addtolength{\abovedisplayskip}{\extraspaces}
\addtolength{\belowdisplayskip}{\extraspaces}
\addtolength{\abovedisplayshortskip}{\extraspace}
\addtolength{\belowdisplayshortskip}{\extraspace}}
\newcommand{\ee}{\end{equation}}

\newcommand{\ba}{\begin{eqnarray}
\addtolength{\abovedisplayskip}{\extraspaces}
\addtolength{\belowdisplayskip}{\extraspaces}
\addtolength{\abovedisplayshortskip}{\extraspace}
\addtolength{\belowdisplayshortskip}{\extraspace}}
\newcommand{\ea}{\end{eqnarray}}

\newcommand{\nonu}{\nonumber \\[.5mm]}
\newcommand{\A}{&\!\!\!}
\newcommand{\Z}{\bf Z}
\newcommand{\R}{\bf R}
\newcommand{\e}{\, {\rm e}}
\newcommand{\bra}[1]{\left\langle {#1} \right\vert}
\newcommand{\ket}[1]{\left\vert {#1} \right\rangle}
\newcommand{\VEV}[1]{\left\langle {#1} \right\rangle}

\newcommand{\newsection}[1]{
\vspace{7mm}
\pagebreak[3]
\addtocounter{section}{1}
\setcounter{equation}{0}
\setcounter{subsection}{0}
\setcounter{footnote}{0}
\begin{center}
{\large {\bf \thesection. #1}}
\end{center}
\nopagebreak
\medskip
\nopagebreak
\hspace{3mm}}

\newcommand{\il}{\lambda_{{}_{{}_{\!\!\!\!\scriptstyle{i}}}}}
\newcommand{\ml}{\lambda_{{}_{{}_{\!\!\!\!\scriptstyle{l}}}}}
\newcommand{\bl}{\lambda_{{}_{{}_{\!\!\!\!\scriptstyle{0}}}}}

\newcommand{\blz}{\lambda_{{}_{{}_{\!\!\!\!\scriptstyle{0}}}}}

\newcommand{\al}{\lambda_{{}_{{}_{\!\!\!\!\scriptstyle{a}}}}}
\newcommand{\cl}{\lambda_{{}_{{}_{\!\!\!\!\scriptstyle{1}}}}}
\newcommand{\dl}{\lambda_{{}_{{}_{\!\!\!\!\scriptstyle{2}}}}}
\newcommand{\el}{\lambda_{{}_{{}_{\!\!\!\!\scriptstyle{3}}}}}
\newcommand{\kl}{\lambda_{{}_{{}_{\!\!\!\!\scriptstyle{4}}}}}

\begin{document}


\thispagestyle{empty}

\begin{flushright}
STUPP--95--141 \\ KITASATO--95--1 \\ August 1995
\end{flushright}
\vspace{.6cm}

\begin{large}

\centerline{{\bf Energy of General Spherically Symmetric}} 

\bigskip

\centerline{{\bf  Solution in Tetrad Theory of Gravitation}} 

\end{large}

\hspace{2cm}


\bigskip

\centerline{
Takeshi SHIRAFUJI$\,{}^\ast$, Gamal G.L. NASHED$\,{}^\ast$
$\!$\footnote[2]{\,Permanent address: Mathematics Department, 
Faculty of Science, Ain Shams University, Cairo, Egypt.} 
and Kenji HAYASHI$\,{}^{\ast\ast}$
}

\bigskip

\centerline{${}^\ast$\ {\it Physics Department, Faculty of Science, 
Saitama University, Urawa 338}}

\bigskip

\centerline{${}^{\ast\ast}$\ {\it Institute of Physics, Kitasato University, 
Sagamihara 228}}


\hspace{2cm}
\\
\\
\\
\\
\\
\\
\\
\\
\\

We find the most general, spherically symmetric solution in a special class
of tetrad theory of gravitation. The tetrad gives the Schwarzschild metric.
The energy is calculated by the superpotential method and by the Euclidean
continuation method. We find that unless the time-space components of the tetrad go to zero faster than ${1/\sqrt{r}}$ at infinity, 
the two methods give 
results different from each other, and  that these results differ from 
the gravitational mass of the
central gravitating body. This fact implies that the time-space components
of the tetrad describing an isolated spherical body must vanish faster than
${1/\sqrt{r}}$ at infinity.

%

\newpage
\newsection{Introduction}

The notion of absolute parallelism was first introduced in physics
by ${\rm Einstein}^{1)}$ trying to unify gravitation and electromagnetism 
into
16 degrees of freedom of the tetrads. His attempt failed, however, because 
there was no Schwarzschild solution in his field equation.

${\rm M\phi ller}^{2)}$ revived the tetrad theory of gravitation and showed 
that a tetrad description of gravitational field allows a more satisfactory
treatment of the energy-momentum complex than that of general relativity. The 
Lagrangian formulation of the theory was given by Pellegrini and ${\rm Plebanski.}^{3)}$ In these attempts the admissible Lagrangians were limited by the 
assumption that the equations determining the metric tensor should coincide
with the Einstein equation. ${\rm M\phi ller}^{4)}$ abandoned this assumption 
and suggested to look for a wider class of Lagrangians, allowing for possible 
deviation from the Einstein equation in the case of strong gravitational 
fields. ${\rm S\acute{a} ez}^{5)}$ generalized M\o ller theory into a scalar
tetrad theory of gravitation. ${\rm Meyer}^{6)}$ showed that M\o ller theory
is a special case of Poincar$\acute{e}$ gauge ${\rm theory.}^{7),8)}$

Quite independently, Hayashi and ${\rm Nakano}^{9)}$ formulated the tetrad 
theory of gravitation as the gauge theory of space-time translation group.
Hayashi and ${\rm Shirafuji}^{10)}$ studied the geometrical and observational 
basis of the tetrad  theory, assuming that the
Lagrangian  be given by a quadratic form of torsion tensor. If invariance
under parity operation is assumed, the most general Lagrangian consists of 
three terms with three unknown parameters to be fixed by experiments, besides a
cosmological term. Two of the three parameters were determined by comparing 
with solar-system ${\rm experiments,}^{10)}$ while only an upper bound has been
estimated for the third ${\rm one.}^{10),11)}$

The numerical values of the two parameters found were very small consistent
with being equal to zero. If these two parameters are equal to zero exactly, 
the theory reduces to the one proposed by
Hayashi and ${\rm Nakano}^{9)}$ and ${\rm M\phi ller,}^{4)}$ which we shall here refer to as the HNM theory for short. This theory differs from 
general
relativity only when the torsion tensor has nonvanishing axial-vector part. It
was also ${\rm shown}^{10)}$ that the Birkhoff theorem can be extended to the 
HNM theory. Namely, for spherically symmetric case in vacuum, which is not 
necessarily time independent, the 
axial-vector part of the torsion tensor should vanish due to the antisymmetric
 part of the field equation, and therefore, with the help of the Birkhoff 
${\rm theorem}^{12)}$ of general relativity we see that the spacetime metric 
is the Schwarzschild.

Mikhail et ${\rm al.}^{13)}$ derived the superpotential of the energy-momentum 
complex in the HNM theory 
and applied it to two  spherically symmetric solutions. It was found that in 
one of the two solutions the gravitational mass does not coincide with the
calculated energy.
 Mikhail et ${\rm al.}^{14)}$ also derived a spherically symmetric solution of 
the HNM theory starting from a tetrad which contains three unknown functions and following Mazumder and ${\rm Ray.}^{15)}$ The solution contains one 
arbitrary function of the radial coordinate {\it r}, and  all previous solutions can be obtained from it. The physical properties of this solution have not yet been examined, however. We show that the underlying metric of their solution 
is just the Schwarzschild metric under certain conditions in consistent with 
the extended Birkhoff theorem mentioned above.

The general form of the tetrad, $\il^\mu$, having spherical
symmetry was given by ${\rm Robertson.}^{16)}$ In the cartesian form it can 
be written as  
\footnote{In this paper Latin indices $(i,j,...)$ represent the vector 
number, and Greek indices $(\mu,\nu,...)$ represent the vector 
components. All indices run from 0 to 3. The spatial part of Latin indices are 
denoted by $(a,b,...)$, while that of greek indices by $(\alpha, \beta,...).$
 In the present convention, latin indices are never raised. The tetrad 
$\il^\mu$ is related to the parallel vector fields ${b_i}^\mu$ of ref.10) 
by $\blz^\mu=i{b_{0}}^\mu$ and $\al^\mu={b_a}^\mu$.} 

\ba
\bl^0 \A= \A iA, \quad \al^0 = C x^a, \quad \bl^\alpha = iD x^\alpha \nonu
\al^\alpha \A= \A \delta_a^\alpha B + F x^a x^\alpha + \epsilon_{a \alpha \beta} S x^\beta, 
\ea
where {\it A}, {\it C}, {\it D}, {\it B}, {\it F}, and {\it S} are functions of {\it t} and $r=(x^\alpha x^\alpha)^{1/2}$, and the zeroth vector $\bl^\mu$ has the factor $i=\sqrt{-1}$ to preserve 
Lorentz signature. We consider an asymptotically flat space-time in this paper, and impose the boundary condition that for $r \rightarrow \infty$
 the tetrad (1・1) approaches the tetrad of Minkowski space-time, 
$\left(\il^\mu \right)= {\rm diag}(i,{\delta_a}^{\alpha})$.

It is the aim of the 
present work to find the most general, asymptotically flat solution with 
spherical symmetry in the HNM theory and calculate the energy of that solution. We do this by two methods and compare the results: one is by applying
 the superpotential of Mikhail et ${\rm al.,}^{13)}$ and the other based on 
the Euclidean continuation method of Gibbons and 
${\rm Hawking.}^{17) \sim 19)}$

In section 2 we briefly review the tetrad theory of 
gravitation. In section 3 we first study the general, spherically symmetric 
solution with a nonvanishing {\it S}-term (see (1・1)), and obtain a solution 
with  one parameter.
Then we study the general, spherically symmetric tetrad without 
the {\it S}-term. All the remaining, unknown functions are allowed 
to depend on {\it t} and {\it r}. We find the general solution with an arbitrary function of {\it t} and {\it r}. We also study the solution of Mikhail et 
${\rm al.}^{14)}$ by transforming it to the  
isotropic, cartesian coordinate. It is then found that their general tetrad is 
just the {\it t}-independent case of our general tetrad without the 
{\it S}-term.
 
In section 4 the energy of the gravitating source is calculated by the 
superpotential method, assuming different asymptotic behaviors for
the unknown function involved in the tetrad.
In section 5 we discuss the Euclidean continuation of the general stationary
tetrad, and calculate the action, the energy and the entropy following the 
 Gibbons-Hawking method.
The final section is devoted to the main results and discussion.

Computer algebra system REDUCE 3.3 is used in some calculations.   

\newsection{The tetrad theory of gravitation}

In this paper we follow M\o ller's ${\rm construction}^{4)}$ 
of the tetrad theory of gravitation based on the 
Weitzenb{\rm $\ddot{o}$}ck space-time. In this 
theory the field variables are the 16 tetrad components $\il^\mu$, from which
the metric is derived by
\be
g^{\mu \nu} \stackrel{\rm def.}{=} \il^\mu \il^\nu.
\ee
The Lagrangian ${\it L}$ is an invariant constructed from
$\gamma_{\mu \nu \rho}$ and $g^{\mu \nu}$, where $\gamma_{\mu \nu \rho}$ is
the contorsion tensor given by
\be
\gamma_{\mu \nu \rho} \stackrel{\rm def.}{=} 
{\il}_{\,{}^{{}^{{}^{\scriptstyle{\mu}}}}}
{\il}_{\,{}^{{}^{{}^{\scriptstyle{\nu;\rho}}}}},
\ee
where the semicolon denotes covariant differentiation with respect to 
Christoffel symbols.
The most general Lagrangian density invariant under parity operation is given by the form 
\be
{\cal L} \stackrel{\rm def.}{=} (-g)^{1/2} \left( \alpha_1 \Phi^\mu \Phi_\mu+
\alpha_2 \gamma^{\mu \nu \rho} \gamma_{\mu \nu \rho}+
\alpha_3 \gamma^{\mu \nu \rho} \gamma_{\rho \nu \mu} \right),
\ee
where
\be
g \stackrel{\rm def.}{=} {\rm det}(g_{\mu \nu})
\ee
and $\Phi_\mu$ is the basic vector field defined by
\be
\Phi_\mu \stackrel{\rm def.}{=} {\gamma^\rho}_{\mu \rho}.
\ee
Here $\alpha_1, \alpha_2,$ and $\alpha_3$ are constants determined by M\o ller 
 such that the theory coincides with general relativity in the weak fields:

\be
\alpha_1=-{1 \over \kappa}, \qquad \alpha_2={\lambda \over \kappa}, \qquad 
\alpha_3={1 \over \kappa}(1-\lambda),
\ee
where $\kappa$ is the Einstein 
constant and  $\lambda$ is a free dimensionless parameter. \footnote{Throughout this paper we use the relativistic units, $c=G=1$ and $\kappa=8\pi$.} The same 
choice of the parameters was also obtained by Hayashi and ${\rm Nakano.}^{9)}$

M\o ller applied the action principle to the Lagrangian density (2・3) and 
obtained
the field equation in the ${\rm form}^{4)}$ 
\be
G_{\mu \nu} +H_{\mu \nu} = -{\kappa} T_{\mu \nu}, 
\ee
\be
F_{\mu \nu}=0,
\ee
where the Einstein tensor $G_{\mu \nu}$ is defined by
\be
G_{\mu \nu}=R_{\mu \nu}-{1 \over 2} g_{\mu \nu} R.
\ee
Here $H_{\mu \nu}$ and $F_{\mu \nu}$ are given by
\be
H_{\mu \nu} \stackrel{\rm def.}{=} \lambda \left[ \gamma_{\rho \sigma \mu}
{\gamma^{\rho \sigma}}_\nu+\gamma_{\rho \sigma \mu} 
{\gamma_\nu}^{\rho \sigma}+\gamma_{\rho \sigma \nu} 
{\gamma_\mu}^{\rho \sigma}+g_{\mu \nu}
\left( \gamma_{\rho \sigma \lambda} \gamma^{\lambda \sigma \rho}-{1 \over 2}
\gamma_{\rho \sigma \lambda} \gamma^{\rho \sigma \lambda} \right) \right]
\ee
and 
\be
F_{\mu \nu} \stackrel{\rm def.}{=} \lambda \left[ \Phi_{\mu,\nu}-\Phi_{\nu,\mu}
-\Phi_\rho \left({\gamma^\rho}_{\mu \nu}-{\gamma^\rho}_{\nu \mu} \right)+
{{\gamma_{\mu \nu}}^{\rho}}_{;\rho} \right],
\ee
and they are symmetric and skew symmetric tensors, 
respectively.

M$\phi$ller assumed that the energy-momentum tensor of matter fields is 
symmetric. In the Hayashi-Nakano theory, however, the energy-momentum tensor
of spin-$1/2$ fundamental particles has nonvanishing antisymmetric part arising
from the effects due to intrinsic spin, and the right-hand side of (2・8) does
not vanish when we take into account the possible effects of intrinsic spin.
Nevertheless, since in this paper we consider only solutions in vacuum, we
refer the tetrad theory of gravitation based on the choice of the parameters,
(2・6), as the Hayashi-Nakano-M$\phi$ller (HNM) theory for short.

It can be ${\rm shown}^{10)}$ that the tensors, $H_{\mu \nu}$ and
 $F_{\mu \nu}$, consist of only those terms which are linear or quadratic 
in the axial-vector part of the torsion tensor, $a_\mu$, defined by 
\be
a_\mu \stackrel{\rm def.}{=} {1 \over 3} \epsilon_{\mu \nu \rho \sigma}
\gamma^{\nu \rho \sigma},
\ee
where $\epsilon_{\mu \nu \rho \sigma}$ is defined by
\be
\epsilon_{\mu \nu \rho \sigma} \stackrel{\rm def.}{=} (-g)^{1/2} 
\delta_{\mu \nu \rho \sigma}
\ee
with $\delta_{\mu \nu \rho \sigma}$ being completely antisymmetric and 
normalized as $\delta_{0123}=-1$.
Therefore, both $H_{\mu \nu}$ and $F_{\mu \nu}$ vanish if the $a_\mu$ is 
vanishing. In other words, when the $a_\mu$ is found to vanish from the
antisymmetric part of the field equations, (2・8), the symmetric part (2・7)
coincides with the Einstein equation.

For the spherically symmetric case which is not necessarily time-independent, 
it was ${\rm shown}^{10)}$ that the antisymmetric part (2・8) implies that the
$a_{\mu}$ should vanish. Then according to the Birkhoff theorm of general 
relativity, the metric of spherically symmetric spacetime in vacuum must be 
the Schwarzschild.

\newsection{Spherically symmetric solutions}

In this section we find the most general, spherically symmetric vacuum
solution of the form (1・1) in the HNM theory. The axial-vector part of the 
torsion tensor, $a^\mu$, is vanishing, and the skew part of the field equation 
is satisfied identically as is explained above. We discuss two cases separately: 
One with $S \neq 0$ and the other with $S=0$. \\

(i) The case with nonvanishing {\it S}-term. 
We start with the tetrad of (1・1) with the six unknown functions of
{\it t} and {\it r}. In order to study the condition that the $a^\mu$ vanishes 
it is 
convenient to start from the general expression for the covariant components of the tetrad,
\ba
{\bl}_{\,{}^{{}^{{}^{\scriptstyle{0}}}}} \A= \A i \check{A}, 
\quad {\al}_{\,{}^{{}^{{}^{\scriptstyle{0}}}}}= \check{C} x^a,
\quad {\bl}_{\,{}^{{}^{{}^{\scriptstyle{\alpha}}}}} = i \check{D} x^\alpha \nonu
{\al}_{\,{}^{{}^{{}^{\scriptstyle{\alpha}}}}}
\A= \A \delta_{a \alpha} \check{B}+\check{F} x^a x^\alpha + 
\epsilon_{a \alpha \beta} \check{S} x^\beta, 
\ea
where the six unknown functions, $\check{A}$, $\check{C}$, $\check{D}$, 
$\check{B}$, $\check{F}$, $\check{S}$, are connected with the six functions of 
(1・1). We can assume without loss of generality that the two functions,
$\check{D}$ and $\check{F}$, are 
vanishing by making use of the freedom to redefine ${\it t}$ and 
${\it r}$. Then the condition that the $a^{\mu}$ vanishes is ${\rm found}^{10)}$ to be. 

\be
 0 = \sqrt{(-g)} a^\mu = \left\{ 
\matrix{
& 2\check{B} \check{S}+ \displaystyle{2 \over 3}r(\check{B} \check{S}'
-\check{B}' \check{S}), \qquad \mu=0, \hfill\cr
& - \left\{ \displaystyle{4 \over 3} \check{C} \check{S}+ 
\displaystyle{2 \over 3}(\check{\dot{B}}\check{S}
-\check{B}\check{\dot{S}}) \right\} x^{\alpha}, \quad \mu={\alpha} \hfill\cr 
}\right.
\ee
with $\check{S}'= {d \check{S}/dr}$  and  
$\check{\dot{S}}= {d \check{S}/dt}$. This condition can be 
solved to give
\be
 \check{C}=0, \quad \check{S}={{\xi} \over r^3} \check{B},
\ee
where $\xi$ is a constant with dimension of $(length)^2$. 

The symmetric part of the field equations now coincides with the Einstein
equation and gives the Schwarzschild metric. The metric tensor formed of
the tetrad (3・1) with (3・3) is not of the isotropic form, however, and the
space-space components involve a term proportional to $x^\alpha x^\beta$. We 
can eliminate such a nondiagonal term of the metric tensor by a scale change of
the space coordinate from ${x^{\alpha}}$ to ${X^{\alpha}}= \left({\rho/r} 
\right){x^{\alpha}}$.
Here and henceforth we denote by $\rho$ the radial variable of the isotropic 
coordinate. After elementary calculation we see that $\rho$ is given by
\be
\rho ={r \over \sqrt{2}} \left(1+\sqrt{1+{\xi^2 \over r^4}}  \right)^{1/2}.
\ee

The metric tensor now takes the well-known isotropic form. Applying the scale 
change to the tetrad (3・1) with (3・3), we finally obtain the general, 
spherically symmetric solution with nonvanishing {\it S}-term: the 
nonvanishing, covariant components of the tetrad are given by
\ba
\hspace*{-1cm}
\!\!{\bl}_{\,{}^{{}^{{}^{\scriptstyle{0}}}}} 
\A=\A i{ \left(1-\displaystyle{m \over 2 \rho} \right) 
 \over \left(1+\displaystyle{m \over 2 \rho} \right)},  \nonu
\hspace*{-1cm}
\!\!{\al}_{\,{}^{{}^{{}^{\scriptstyle{\alpha}}}}} 
\A=\A \left(1+ \displaystyle{m \over 2 \rho} \right)^2 
\left[\sqrt{1-(r\check{S})^2} \delta_{a \alpha} + 
\left(1-\sqrt{1-(r\check{S})^2} \right) {X^a X^\alpha \over {\rho}^2}
+(r\check{S}) \epsilon_{a \alpha \beta} {X^\beta \over \rho} \right]\!,
\ea
where $\check{S}$ is given by
\be
\check{S}= \displaystyle{{\xi} \over \left(1+ 
\displaystyle {{\xi}^2 \over 4r^4} \right) r^3}.
\ee

(ii) The case without the {\it S}-term.
In this case the axial-vector part of the torsion tensor is identically 
vanishing. Thus, when
this tetrad is applied to the field equations, the 
skew part is automatically satisfied in vacuum and the solution of the symmetric part is the Schwarzschild. Therefore, the solution of the 
form (1・1) with $S=0$ can be obtained from the diagonal tetrad of the 
Schwarzschild metric by a {\it local Lorentz transformation} which keeps 
spherical symmetry, 
\be
\left(\Lambda_{k l} \right) = 
\left(
\matrix{
\sqrt{H^2+1} & i H \displaystyle{X^b \over \rho}  \vspace{3mm} \cr 
-i H \displaystyle{X^a \over \rho} & {\delta_a}^b+\left(\sqrt{H^2+1}-1 \right)
\displaystyle{X^aX^b \over {\rho}^2} 
\cr}\right),
\ee
where ${\it H}$ is an arbitrary function of {\it t} and $\rho$. Namely, we see 
that 
\be
\il^\mu= \Lambda_{i l} \ml^{(\small 0) \mu}
\ee
is the most general, spherically symmetric solution without the 
{\it S}-term. Here $\ml^{ (\small 0) \mu}$ is the diagonal tetrad in the 
isotropic, cartesian coordinate given by

\be
\bl^{(\small 0) 0} = i {\left(1+\displaystyle{m \over 2\rho} \right) \over
\left(1-\displaystyle{m \over 2\rho} \right)} , \qquad
\al^{( \small 0) \alpha} = {{\delta_a}^\alpha \over \left(1+ 
\displaystyle{m \over 2\rho} \right)^2}. 
\ee
The explicit form of the ${\il}^\mu$ is then given by

\ba
\bl^0 \A=\A i{(1+\displaystyle{m \over 2 \rho}) 
 \over (1-\displaystyle{m \over 2 \rho})} \sqrt{H^2+1}, \nonu
\bl^\alpha \A=\A i{H \over (1+\displaystyle{m \over 2 \rho})^2} 
{X^\alpha \over \rho}, \nonu
\al^0 \A=\A {H (1+\displaystyle{m \over 2\rho}) \over 
(1-\displaystyle{m \over 2 \rho})} {X^a \over \rho}, \nonu
\al^\alpha \A=\A 
{1 \over (1+ \displaystyle{m \over 2\rho})^2} 
\left[{\delta_a}^\alpha + (\sqrt{H^2+1}-1){X^a X^\alpha \over {\rho}^2} \right].
\ea
It is clear that if $\xi$ and $H(r,t)$ are equal to zero the two classes of  
solutions given by \\ 
(3・5) and (3・10) coincide with each other, and reduce to 
the solution given by Hayashi and ${\rm Shirafuji}^{10)}$ in the special case 
$p=q=2$. 

Now let us compare the solution (3・10) with that given by
 Mikhail et ${\rm al.}^{14)}$ They started from a spherically symmetric tetrad 
with three unknown functions, which is given in the spherical polar coordinate 
by

\be
\left(\il^\mu \right)= 
\left(
\matrix{
iA & iDr & 0 & 0 \vspace{3mm} \cr 
0 & B \sin\theta \cos\phi & \displaystyle{B \over r}\cos\theta \cos\phi 
 & -\displaystyle{B \sin\phi \over r \sin\theta} \vspace{3mm} \cr
0 & B \sin\theta \sin\phi & \displaystyle{B \over r}\cos\theta \sin\phi 
 & \displaystyle{B \cos\phi \over r \sin\theta} \vspace{3mm} \cr
0 & B \cos\theta & -\displaystyle{B \over r}\sin\theta  & 0 \cr
}
\right)
\ee
and they applied it to the field equations, (2・7) and (2・8), to obtain a
solution of the form 
\be
A=\displaystyle{K_1 \over 1- \displaystyle {rB' \over B} }, \quad
D^2={1 \over \left(1- \displaystyle{rB' \over B} \right)^2 } 
\left({B \over r } \right)^3 
\biggl[K_2+{rB' \over B}({rB' \over B}-2) {r \over B} \biggr],
\ee
where $K_1$ and $K_2$ are constants of integration and {\it B} is an arbitrary 
function of {\it r}. The line-element squared takes the form
\be
ds^2=-{(B^2-D^2 r^2) \over A^2B^2} dt^2-{2Dr \over AB^2}dr dt+ {1 \over B^2}
(dr^2+r^2 d\Omega^2)
\ee
with ${d\Omega^2=d\theta^2+\sin^2\theta d\phi^2}$. We assume $B(r)$ to be 
nonvanishing so that the surface area of the sphere of a constant ${\it r}$
be finite. We also assume that
$A(r)$ and $B(r)$ satisfy the asymptotic condition,
 $\lim_{r \to \infty} A(r)$=$\lim_{r \to \infty} B(r)=1$ and 
 $\lim_{r \to \infty} rB'=0$. Then, we can show from  (3・12) and (3・13) that
(1) $K_1=1$, (2)\, $B(r)>0$, (3) \, $\lim_{r \to \infty} rD(r)=0,$ and
(4) if $B-r B'$ vanishes at some point, then $1-B K_2/r<0$ at that point.

Using the coordinate transformation \\
\be
dT=dt+{ADr \over B^2-D^2r^2}dr,  
\ee
we can eliminate the cross term of (3・13) to obtain
\ba
ds^2= -\eta_1 dT^2 +{1 \over \eta_1} {dr^2 \over A^2B^2} +
{r^2 \over B^2}d\Omega^2
\ea
with $\eta_1={(B^2-D^2 r^2)/A^2B^2}$.
Taking the new radial coordinate $R={r/B}$, we finally get
\be
ds^2= -\eta_1 dT^2 +{dR^2 \over \eta_1} +R^2d\Omega^2,
\ee
where 
\be
\eta_1(R)=(1-{K_2 \over R}).
\ee 
Then, (3・16) coincides with the 
Schwarzschild metric with the mass, $m= {K_2/2}$, and hence the 
solution in the case of the spherically symmetric tetrad gives no more 
than the Schwarzschild solution when $1-{rB'/B}$ has no zero and ${\it R}$ is 
monotonically increasing function of ${\it r}$.
If $1-{rB'/B}$ has zeroes, the line-element (3・13)
is singular at these zeroes 
which lie inside the event horizon as is seen from the property (4)
mentioned above. We shall study in the future whether this singularity at
zero-points of $1-{rB'/B}$ is physically acceptable or not.

The tetrad (3・11) has been subject to two steps of coordinate 
transformations from $(t, r, \theta, \phi)$ to $(T, R, \theta, \phi)$. We now
make a further transformation from $( T, R, \theta, \phi)$ to the 
isotropic coordinate $(T, X^{\alpha})$ with  ${\alpha}=1,2$ and $3$, where the 
line-element squared takes the well-known form 
\be
ds^2= -{ \left(1-\displaystyle{m \over 2\rho} \right)^2 
 \over \left(1+\displaystyle{m \over 2\rho} \right)^2}dT^2 
+\left(1+\displaystyle{m \over 2\rho} \right)^4 (dX^{\alpha})^2
\ee
with $\rho=(X^\alpha X^\alpha)^{1/2}$. After lengthy calculation the tetrad is expressed by
\ba
\bl^0 \A=\A i{ \left(1+\displaystyle{m \over 2\rho} \right)^2 
 \over \left(1-\displaystyle{m \over 2\rho} \right)^2}
 \left[ 1-\displaystyle{\rho B' 
 \over B^3 \left(1+\displaystyle{m \over 2\rho} \right)^4} \right], \nonu
\bl^{\alpha} \A=\A  {\displaystyle 2i \sqrt{m \over 2\rho} \over 
 \left(1 -\displaystyle{m \over 2\rho} \right)
 \left(1 +\displaystyle{m \over 2\rho} \right)^2 } 
 \left[1- \displaystyle{{\rho}^2 B' \over m B^3 
 \left(1 +\displaystyle{m \over 2\rho} \right)^2} 
 + {{\rho}^3 B'^2 \over 2m B^6 \left(1+ \displaystyle{m \over 2\rho}\right)^6} 
 \right]^{1/2} \displaystyle{X^{\alpha} \over \rho}, \nonu
{\al}^0 \A=\A  {\displaystyle 2 \sqrt{m \over 2\rho}
\left(1+ \displaystyle{m \over 2\rho} \right) \over 
 \displaystyle \left(1- \displaystyle{m \over 2\rho} \right)^2}
 \left[1- \displaystyle{{\rho}^2 B' \over m B^3 
\left(1+ \displaystyle{m \over 2\rho} \right)^2}
+ \displaystyle{{\rho}^3 B'^2 \over 2m B^6 \left(1+ \displaystyle
{m \over 2\rho} \right)^6} \right]^{1/2} \displaystyle{X^a \over \rho}, \nonu
{\al}^{\alpha} \A=\A {1 \over \left(1+ \displaystyle{m \over 2\rho} \right)^2} 
\left[{\delta_a}^b + \displaystyle{\displaystyle {m \over \rho} \over 
\left(1- \displaystyle{m \over 2\rho} \right)}
\left(1- \displaystyle{{\rho}^2 B' \over m B^3 
\left(1+ \displaystyle{m \over 2\rho} \right)^3} 
\right){X^aX^{\alpha} \over {\rho}^2} \right].
\ea
It is easy to verify that this tetrad can be obtained from (3・10) by choosing 
the function {\it H} as
\be
H= 2{ \displaystyle \sqrt{m \over 2\rho} \over \left(1- 
\displaystyle{m \over 2\rho} \right)} 
\left[1-{{\rho}^2B' \over mB^3 \left(1+ \displaystyle{m \over 2\rho} \right)^2}
+{{\rho}^3B'^2 \over 2mB^6 \left(1+ 
\displaystyle{m \over 2\rho} \right)^6} \right]^{1/2}.
\ee

\newpage

\newsection{Energy with M\o ller's method}

The superpotential of the HNM theory is given by Mikhail et ${\rm al.}^{13)}$ 
as
\be
{{\cal U}_\mu}^{\nu \lambda} ={(-g)^{1/2} \over 2 \kappa}
{P_{\chi \rho \sigma}}^{\tau \nu \lambda}
\left[\Phi^\rho g^{\sigma \chi} g_{\mu \tau} 
 -\lambda g_{\tau \mu} \gamma^{\chi \rho \sigma}
-(1-2 \lambda) g_{\tau \mu} \gamma^{\sigma \rho \chi}\right], 
\ee
where ${P_{\chi \rho \sigma}}^{\tau \nu \lambda}$ is 
\be
{P_{\chi \rho \sigma}}^{\tau \nu \lambda} \stackrel{\rm def.}{=} 
{{\delta}_\chi}^\tau {g_{\rho \sigma}}^{\nu \lambda}+
{{\delta}_\rho}^\tau {g_{\sigma \chi}}^{\nu \lambda}-
{{\delta}_\sigma}^\tau {g_{\chi \rho}}^{\nu \lambda}
\ee
with ${g_{\rho \sigma}}^{\nu \lambda}$ being a tensor defined by
\be
{g_{\rho \sigma}}^{\nu \lambda} \stackrel{\rm def.}{=} 
{\delta_\rho}^\nu {\delta_\sigma}^\lambda-
{\delta_\sigma}^\nu {\delta_\rho}^\lambda.
\ee
The energy is expressed by the surface ${\rm integral,}^{20)}$
\be
E=\lim_{\rho \rightarrow \infty}\int_{\rho=constant} {{\cal U}_0}^{0 \alpha} 
n_\alpha dS, 
\ee
where $n_\alpha$ is the unit 3-vector normal to the surface element ${\it dS}$.

 Firstly, let us consider the case with vanishing {\it S}-term, for which the 
tetrad (3・10) takes the following form asymptotically:

\ba
\bl^0 \A=\A i \left[1+ \left({m \over \rho} +\displaystyle{H^2 \over 2}
 \right) \right] , \nonu
\bl^\alpha \A=\A i H \displaystyle{X^\alpha \over \rho}, \nonu
\al^0 \A=\A H \displaystyle{X^a \over \rho}, \nonu
\al^\alpha \A=\A \left(1- \displaystyle{m \over \rho} \right){\delta_a}^\alpha 
+ \displaystyle{H^2 \over 2} \displaystyle{X^a X^b \over {\rho}^2},
\ea
where we understand that {\it H} denotes the leading term of the function 
$H(\rho,t)$ for $\rho \rightarrow \infty$.
We discuss three different cases separately according to the asymptotic
form of {\it H}. \\

({\bf i}) The case with $H \sim {f(t)/\sqrt{{\rho}^{1-\epsilon}}}$ for a 
constant $\epsilon$ satisfying $1 > \epsilon >0$. The superpotential of (4・1) 
behaves for large $\rho$ as 

\be
{{\cal U}_0}^{0 {\alpha}} =-{f^2 \over \kappa {\rho}^2} 
 {X^{\alpha} \over \rho} {\rho}^{\epsilon}.
\ee
Substituting (4・6) into (4・4), we see that the integral (4・4) is 
divergent. Thus, this case should be rejected from our consideration. \\

({\bf ii}) The case with $H \sim {f(t)/\sqrt{\rho}}$.
The superpotential of (4・1) behaves like
\be
{{\cal U}_0}^{0 {\alpha}} ={2m+f^2 \over \kappa \rho^2} 
{X^{\alpha} \over {\rho}}   
\ee
in this case. Substituting (4・7) into (4・4), we get 
\be
E=m+\displaystyle{f^2 \over 2}.
\ee
Therefore, if  $f \neq 0$ the value of the energy differs from the gravitational mass {\it m}. \\

({\bf iii}) The case with $H \sim {f(t)/\sqrt{{\rho}^{1+\epsilon}}}$ 
for a positive constant $\epsilon$. 
The superpotential of (4・1) behaves like
\be
{{\cal U}_0}^{0 {\alpha}} = {2m \over \kappa \rho^2} {X^{\alpha} \over {\rho}} 
\ee
for this case. Calculating the energy from (4・4), we get 
\be
E= m
\ee
in agreement with the gravitational mass.

Now let us consider the solution with non-vanishing {\it S}-term. 
The asymptotic behavior of the tetrad (3・5) is given by (4・5) with vanishing
{\it H}. Therefore, the energy is given by (4・10) also in this case.

\newsection{Thermal properties of the spherically 

\quad\ symmetric solution}

Gibbons and ${\rm Hawking}^{17) \sim 19)}$ discussed the thermal properties of 
the Schwarzschild solution, for which the line-element squared takes the 
positive-definite standard form
\be
ds^2= + \left(1-{2m \over r} \right) d{\tau}^2+ \left(1-{2m \over r} 
\right)^{-1} dr^2+ r^2 d{\Omega^2},
\ee
after the Euclidean continuation of the time variable, $t= -i\tau$. 
By the transformation $x=4m \left(1- {2m/r} \right)^{1/2}$, the line-element
 squared becomes 
\be
ds^2= +\left({x \over 4m} \right)^2 d {\tau}^2+ \left({r^2 \over 4m^2} \right)^2 dx^2+ r^2 d{\Omega^2},
\ee
which shows that $\tau$ can be regarded as an angular variable with period
$8{\pi}m$. Now the Euclidean section of the Schwarzschild solution is the region defined by $8{\pi}m \ge \tau \ge 0$ and $x>0$, where the metric is positive 
definite, asymptotically flat and non-singular. They calculated the Euclidean 
action, ${\hat I},$ of 
general relativity from the surface term as follows:
\be
{\hat I}=4{\pi}m^2={\beta^2 \over 16\pi},
\ee
where $\beta=8{\pi}m =T^{-1}$ with {\it T} being interpreted as the absolute 
temperature of the Schwarzschild black hole.

For a canonical ensemble the energy is given by

\be
E={{\sum_{n} E_n e^{-\beta E_n}} \over {\sum_{n} e^{-\beta E_n}}}=
 -{\partial \over \partial \beta} {\rm log} Z,
\ee
where $E_n$ is the energy in the ${\it n}$th state, and {\it Z} is the 
partition function, which is in the tree approximation related to the Euclidean action of the classical solution by
\be
{\hat I}=-{\rm log} Z.
\ee
Use of (5・3) and (5・5) in (5・4) gives
\be
E={\beta \over 8\pi}=m.
\ee
They also calculated the entropy of the Schwarzschild black hole to obtain 

\be
S=-\sum_{n} P_n {\rm log}P_n = \beta E +{\rm log}Z=4{\pi}m^2={1 \over 4}A,
\ee
where $P_n =Z^{-1} e^{-\beta E_n}$, and {\it A} is the area of the event horizon of the Schwarzschild black hole.

Let us apply the above procedure of Gibbons and Hawking to our spherically 
symmetric, stationary solutions, namely, to the tetrad (3・5) with the {\it S}
-term and to the tetrad (3・8) without the {\it S}-term, where the arbitrary 
function {\it H} is assumed to be independent of {\it t}. Since these solutions give the 
Shwarzschild metric, the variable $\tau$ can be regarded as an angular variable
with period $8 \pi m$ after the Eculidean continuation, $t=-i \tau$. The 
Euclidean continuation of the diagonal tetrad in the isotropic, 
cartesian coordinate is given by
\be
{\kl}_{\,{}^{{}^{{}^{\scriptstyle{4}}}}}^{(0)} 
={{1-\displaystyle{m \over 2\rho}}  \over 
{1+\displaystyle{m \over 2\rho}}}, 
\qquad {\al}_{\,{}^{{}^{{}^{\scriptstyle{\alpha}}}} 
}^{(0)} =\left(1+{m \over 2\rho} \right)^2 
\delta_{a \alpha},
\ee
where we use the index 4 instead of 0 in the Euclidean section. The Euclidean
 continuation of the general tetrad with vanishing {\it S}-term is then obtained from this diagonal one by local {\it SO}(4) rotations preserving spherical 
symmetry,
\be
\left(\Lambda_{k l}  \right) = 
\left(
\matrix{
\sqrt{1-{\hat H}^2} & {\hat H} \displaystyle{X^b \over \rho}  \vspace{3mm} \cr 
-{\hat H} \displaystyle{X^a \over \rho}  & \delta_{a b}+
\left(\sqrt{1-{\hat H}^2}-1 \right)
\displaystyle{X^aX^b \over {\rho}^2} \cr
}
\right)
\ee
where the indices {\it k} and {\it l} run over 4,1,2 and 3 in the Euclidean 
section, and ${\hat H}$ is an arbitrary function of ${\rho}$. Here the matrix
 (5・9) is related to the local Lorentz transformation matrix (3・7) by the
continuation, $H=-i{\hat H}$.

As for the solution with nonvanishing {\it S}-term, its Euclidean 
continuation is obtained simply by removing the factor {\it i} from the first 
equation of (3・5).

For these solutions in the Euclidean section, the axial-vector part of the
torsion tensor vanishes, and the Euclidean action is given by

\be
{\hat I}= -{1 \over 2\kappa} \int \sqrt{g} \left(R-2 {\Phi^{\mu}}_{;\mu} \right) d^4x= {1 \over \kappa} \int \sqrt{g} {\Phi^{\mu}}_{;\mu} d^4x.
\ee
where {\it R} is the Riemann-Christoffel scalar curvature, and $\Phi^\mu$ is 
the basic vector field defined by (2・5).
As in the previous section, we divide the solutions in the Euclidean  section
into three cases according to the asymptotic behavior of the arbitrary
function ${\hat H}(\rho)$ in (5・9). \\

({\bf i}) The case with $ {\hat H}(\rho) \sim {{\hat f}/\sqrt{{\rho}^
{1-\epsilon}}}$ for $1>\epsilon>0$ \footnote{Here ${\hat f}$ is a constant in 
contrast with the $f(t)$ introduced in the previous section.}. 
The action of (5・10) is divergent, 
which also justifies our conclusion of the previous section that this case 
should be rejected. \\

({\bf ii}) The case with ${\hat H}(\rho)\sim {{\hat f}/\sqrt{\rho}}$. 
The surface integral (5・10) is calculated to give
\be
{\hat I}=4{\pi}m(m-{\hat f}^2),
\ee
and the use of this in (5・4) gives the energy
\be
E=m-{\partial \over \partial m}({m{\hat f}^2 \over 2}) .
\ee
We notice that this value of energy is different from that given by using the 
superpotential in section 4 and also from that of general relativity.
The entropy is obtained from the second equation of (5・7) as
\be
S=4{\pi}m^2(1-{\partial {\hat f}^2 \over \partial m}).
\ee

({\bf iii})  The case with ${\hat H}(\rho)\sim {{\hat f}/\sqrt{{\rho}^
{1+\epsilon}}}$ for $\epsilon>0$. The Euclidean action is given by
\be
{\hat I}=4{\pi}m^2
\ee
 in this case, and the energy is
\be
E=m.
\ee
This value is the same as that given in section 4, and the value of the entropy
is
\be
S=4{\pi}m^2={1 \over 4} A,
\ee
showing that both the energy and the entropy are in agreement with those of 
general relativity.

Finally, we notice that the same result as the case (iii) is obtained also for 
the solution with nonvanishing {\it S}-term.

\newsection{Main results and discussion}

 In this paper we have studied the most general, spherically symmetric solutions in the HNM tetrad theory of gravity. According to the Birkhoff theorem of this ${\rm theory,}^{10)}$
 the axial-vector part of the torsion tensor, $a^\mu$, 
should vanish for any spherically
symmetric solution, and accordingly the underlying spacetime metric must be
the Schwarzschild.

Tetrads with spherical symmetry are classified into two groups according as 
whether the space-space components ,$\al^\alpha$, have the {\it S}-term,
namely the term of $S(t,\rho) \epsilon_{ a \alpha \beta} x^\beta$,or not. When
 the {\it S}-term is {\it non-vanishing}, the tetrad is severely restricted by 
the condition that the $a^\mu$ be vanishing, and
we get a family of solutions with a constant parameter. On the other hand, when
 the {\it S}-term is {\it vanishing}, the $a^\mu$ identically vanishes, and 
accordingly we get a family of solutions with an arbitrary function of t and 
$\rho$, and establish its relation with the solution of 
Mikhail et ${\rm al.}^{14)}$
 
We have applied the superpotential given by Mikhail et ${\rm al.}^{13)}$ 
to calculate the energy of the central gravitating body.
As for the tetrad without the {\it S}-term, we discuss three cases 
separately according to the asymptotic behavior of $\bl^\alpha(t,\rho)$.
(i) When $\bl^\alpha \sim {\rho}^{-(1-\epsilon)/2}$
for $1> \epsilon >0$,
 we find that the energy is divergent. So we reject this case from our 
consideration. (ii) When $\bl^\alpha \sim f(t) {\rho}^{-1/2}$,
the energy is given by $E=m+f^2/2$. This case has many problems, however.
 Firstly the
energy differs from the gravitational mass {\it m}, and secondly the energy 
now depends on time, because $f(t)$ is a function of {\it t} 
in general. Is it physically acceptable? We will leave this problem
to another paper, but our preliminary investigation suggests that the answer 
will be negative, because the equivelance between the gravitational mass
 and the inertial mass is violated. (iii) When $\bl^\alpha 
\sim {\rho}^{-(1+\epsilon)/2}$ for $\epsilon >0$, the energy agrees with {\it m}, and this case is very satisfactory. We also find that the tetrad 
with the {\it S}-term gives the same result as in case (iii).

We have then used the method of Gibbons and Hawking to calculate the energy
 for the stationary solutions.
We classify the tetrad without the {\it S}-term in the Euclidean section into 
three cases 
according to the asymptotic behavior of ${\kl}^{\alpha}(\rho)$. 
(i) When $\kl^\alpha \sim {\rho}^{-(1-\epsilon)/2}$
for $1> \epsilon >0$, the action in the Euclidean section diverges, and 
hence this case must be rejected. (ii) When $\kl^\alpha \sim {\rho}^{-1/2}$
the calculated energy differs both from the gravitational mass and from
the value obtained by the superpotential method. (iii) Finally 
when $\kl^\alpha \sim {\rho}^{-(1+\epsilon)/2}$
for $\epsilon >0$, the energy is found to coincide with the gravitational mass.
As for the tetrad with the {\it S}-term, the energy agrees with that of case 
(iii).

Is there any inconsistency between the two methods to calculate the energy? 
If so, which is the correct formula for calculating energy, that given
by M\o ller or that by Gibbons and Hawking? If the two methods are to be
consistent, our result implies that we must reject the case (ii), requiring that $\bl^\alpha$ (or $\kl^\alpha$ in the Euclidean section) vanishes faster than 
$1/\sqrt{\rho}$ at infinity.

Finally we make a brief comment concerning the geometrical meaning of the 
above result. The spherically symmetric, vacuum solutions discussed in this
paper have the following property in common: They give the Schwarzschild metric, and
define the axial-vector part of the torsion tensor $a^\mu$ which is identically
vanishing. This means that these solutions are indistinguishable from each
other observationally, as far as one uses the photons or spin-$1/2$ fundamental
particles as test particles to explore the underlying
structure of spacetime. In this sense these
solutions are all physically equivalent with each other. Geometrically speaking, any one of these solutions can be chosen as parallel vector fields to define
extended absolute parallelism, and then the underlying spacetime becomes an 
extended
Weizenb$\ddot{o}$ck ${\rm spacetime}^{10)}$ \footnote{For more details of 
extended Weizenb$\ddot{o}$ck spacetime see section 8 of ref. 10).}
, in which local Lorentz 
transformations preserving the condition, $a^\mu=0$, are allowed. According to 
the above result, allowed local Lorentz transformations must also preserve the
required asymptotic behavior,  $\bl^\alpha \sim {\rho}^{-(1+\epsilon)/2}$ for 
$\epsilon>0$.

We have summarized these results in Table 1.

\bigskip
\bigskip

\centerline{\Large{\bf Acknowledgements}}

One of the authors (G.N.) would like to thank Japanese Government for 
supporting him with Monbusho Scholarship and also want to express his deep 
gratitude to all the members of Physics Department at Saitama University, 
especially to Prof.\ Saso and Dr.\ Yamaguchi. 
\bigskip


\newpage

\centerline{\Large{\bf References}}

\bigskip

\begin{enumerate}

\item[{1)}] A. Einstein, 
 {\it Sitzungsber.\ Preuss.\ Akad.\ Wiss.\ } (1928), 217. 

\item[{2)}] C. M\o ller,
 {\it Ann. of Phys. } {\bf 12} (1961), 118. 

\item[{3)}] C. Pellegrini and J. Plebanski, 
 {\it Mat.\ Fys.\ Skr.\ Dan.\ Vid.\ Selsk.\ } {\bf 2} (1963), 4. 

\item[{4)}] C. M\o ller, 
 {\it Mat.\ Fys.\ Medd.\ Dan.\ Vid.\ Selsk.\ } {\bf 39} (1978), 13. 

\item[{5)}] D. S$\acute{a}$ez, 
 {\it Phys.\ Rev.\ } {\bf D27} (1983), 2839. 

\item[{6)}] H. Meyer,  
 {\it Gen.\ Rel.\ Grav.\ }{\bf 14} (1982), 531. 

\item[{7)}] K. Hayashi and T. Shirafuji, 
 {\it Prog.\ Theor.\ Phys.\ }{\bf 64}, 866, 883, 1435, 2222; 
{\bf 65} (1980), 525. 

\item[{8)}] F.W. Hehl, J. Nitsch and P. von der Heyde,  
 in {\it ''General Relativity and Gravitation''}, A.\ Held, ed. (Plenum
Press, New York, 1980).  

\item[{9)}] K. Hayashi and T. Nakano,  
{\it Prog.\ Theor.\ Phys.\ }{\bf 38} (1967), 491. 

\item[{10)}] K. Hayashi and T. Shirafuji,  
{\it Phys.\ Rev.\ }{\bf D19} (1979), 3524. 

\item[{11)}] S. Miyamoto and T. Nakano,  
{\it Prog.\ Theor.\ Phys.\ }{\bf 45} (1971), 295. 

\item[{12)}] G. Birkhoff, {\it ''Relativity and Modern Physics''} 
(Harvard Univ. Press, Cambridge, 1923). 

\item[{13)}] F.I. Mikhail,  M.I. Wanas, A. Hindawi and E.I. Lashin,  
{\it Int.\ J.\ of Theor.\ Phys.\ }{\bf 32} (1993), 1627. 

\item[{14)}] F.I. Mikhail, M.I. Wanas, E.I. Lashin and A. Hindawi,  
{\it Gen.\ Rel.\ Grav.\ }{\bf 26} (1994), 869. 

\item[{15)}] A. Mazumder  and D. Ray, 
{\it Int.\ J.\ of Theor.\ Phys.\ }{\bf 29} (1990), 431. 

\item[{16)}] H.P. Robertson, 
 {\it Ann.\ of Math.\ (Princeton)} {\bf 33} (1932), 496. 

\item[{17)}] G.W. Gibbons and S.W. Hawking,  
{\it Phys.\ Rev.\ }{\bf D15} (1977), 2752. 

\item[{18)}] S.W. Hawking, 
{\it Phys.\ Rev.\ }{\bf D14} (1976), 2460. 

\item[{19)}] S.W. Hawking,  
 in {\it ''General Relativity''}, S. W. Hawking and W. Israel,
eds. (Cambridge Univ. Press, Cambridge, 1979). 

\item[{20)}] C. M\o ller,
{\it Ann. of Phys. } {\bf 4} (1958), 347.

\end{enumerate}
\newpage

Table 1. Summary of the main results. The general solution with spherical
symmetry is classified into two groups according as the space-space components,
$\al^\alpha$, have the term, $S \epsilon_{a \alpha \beta} x^\beta$
(referred to as the {\it S}-term for short), or not. The general solution 
without
the {\it S}-term has an arbitrary function, so it is further classified
into three classes according to the asymptotic behavior of $\bl^\alpha$ (or
$\kl^\alpha$ in the Euclidean section): (i) $\bl^\alpha$ 
(or $\kl^\alpha$) $\sim {\rho}^{-(1-\epsilon)/2}$ for $1>\epsilon>0$,
(ii) $\bl^\alpha$ $\sim f{\rho}^{-1/2}$
(or $\kl^\alpha \sim {\hat f}{\rho}^{-1/2})$, and 
(iii) $\bl^\alpha$ 
(or $\kl^\alpha$) $\sim {\rho}^{-(1+\epsilon)/2}$ for $\epsilon>0$. The general
solution with the {\it S}-term has a constant parameter, and its components,
 $\bl^\alpha$ (or $\kl^\alpha$ in the Euclidean section), are vanishing.


\end{document}